\documentclass[aps, prd, fleqn, nofootinbib]{revtex4-2}

\usepackage{
	amsmath, 
	amssymb, 
	graphicx,
	hyperref, 
	float, 
	enumerate, 
	pgffor,
    longtable,
	multirow,
	mathrsfs,
	bm,
	slashed,
	xcolor,
    lineno,
    soul,
    pdfpages,
    breqn,
    courier,
    xparse,
    fvextra
}

\usepackage[export]{adjustbox}

\hypersetup{
	colorlinks=true
}

\usepackage{tikz}

\def\bb#1{\left[#1\right]}

\def\CC#1{\Big\{#1\Big\}}

\def\nn{\nonumber \\ &}

\def\d#1{\mathop{{\rm d}#1}}

\def\vec#1{\bm{#1}}

\allowdisplaybreaks

\def\GeV{{\rm GeV}}
\def\TeV{{\rm TeV}}

\DeclareMathAlphabet{\mymathbb}{U}{BOONDOX-ds}{m}{n}

\def\bar{\overline}
\def\tilde{\widetilde}

\DefineVerbatimEnvironment{verbatim}{Verbatim}{breaklines=true}

\def\mll{{m_{\ell\ell}}{}}

\def\pt{{p_T}{}}

\def\OMIT#1{}

\usepackage{listings}
\lstdefinelanguage{mypython}{
  language=Python,
  keywords={def, return, if, elif, else, for, while, in, import, from, as, with, try, except, raise, class, self, None, True, False},
  keywordstyle=\color{blue}\bfseries,
  ndkeywords={int, float, str, list, dict, set, tuple},
  ndkeywordstyle=\color{magenta},
  identifierstyle=\color{black},
  comment=[l]{\#},
  commentstyle=\color{gray}\ttfamily,
  stringstyle=\color{orange},
  sensitive=true
}
\makeatletter
\AtBeginDocument{\let\LS@rot\@undefined}
\makeatother 

\begin{document}

\title{Naive $T$-odd Drell-Yan angular coefficients as a probe of the dimension-8 SMEFT}
\author{Frank Petriello${}^{1}$, Kaan \c{S}im\c{s}ek${}^{2}$ \\
$ $ \\
${}^1$ \textsl{Department of Physics \& Astronomy, Northwestern University, Evanston, Illinois 60208, USA} \\
${}^2$ \textsl{Department of Physics, Kennesaw State University, Kennesaw, Georgia 30144, USA}}

\begin{abstract}

    We propose the ``naive" $T$-odd Collins-Soper moments in the Drell-Yan process as probes of previously unexplored directions in the Standard Model Effective Field Theory (SMEFT) parameter space. We show that the moments $A_6$ and $A_7$ in the high invariant mass and transverse momentum region are sensitive to dimension-8 $CP$-odd semi-leptonic four-fermion operators with an additional gluon field strength tensor. Using the projected integrated luminosity of a future high-luminosity LHC, we show that effective ultraviolet scales in the few-TeV range can be probed with this process.

\end{abstract}

\maketitle

\section{Introduction}

The study of the angular distribution of leptons produced in the Drell-Yan (DY) process has long been a focus of hadron-collider studies. The dominant production mechanism is through $s$-channel exchange of a photon or $Z$-boson, indicating that the differential cross section can be written as an expansion in spherical harmonics that truncates at $l=2$:
\begin{align}
\frac{\d \sigma}{\d {m_{\ell \ell}{}^2} \d y \d {\Omega_\ell}} &= \frac{3}{16\pi} \frac{\d\sigma}{\d{m_{\ell\ell}{}^2} \d y}
\CC{
1+\cos(\theta)^2
+\frac{A_0}{2}[1-3\cos(\theta)^2]
+A_1 \sin(2\theta) \cos(\varphi)
+\frac{A_2}{2}\sin(\theta)^2 \cos(2\varphi)
\nn
+A_3 \sin(\theta) \cos(\varphi)
+A_4 \cos(\theta)
+A_5 \sin(\theta)^2 \sin(2\varphi) 
+A_6 \sin(2\theta) \sin(\varphi) 
+A_7 \sin(\theta) \sin(\varphi)
}.
     \label{eq:oldCSexp}
\end{align}
This expansion is valid to all orders in the perturbative QCD expansion and is only weakly broken by higher-order electroweak effects~\cite{Alioli:2020kez}. The $A_i$ coefficients are sensitive to numerous phenomena in QCD, and in the Standard Model (SM) more generally. The Lam-Tung relation $A_0=A_2$ is valid through ${\cal O}(\alpha_s)$ and probes the spin-1/2 nature of the quarks appearing in the initial state~\cite{Lam:1978pu}. The coefficient $A_4$ measures parity violation in the electroweak sector of the SM and enables the extraction of the weak mixing angle at hadron colliders~\cite{CMS:2024ony}. Early studies of parity-nonconserving asymmetries in $W$-boson production at large transverse momentum can be found in~\cite{Hagiwara:1984hi}. Measurements of the $A_i$ coefficients at the LHC test models of new physics~\cite{Alioli:2020kez,Li:2022rag}. Studies have shown the need to extend this basis in the presence of higher-dimensional operators~\cite{Alioli:2020kez}, and have also shown the breaking of the Lam-Tung relation in the presence of electroweak dipole operators~\cite{Gauld:2024glt,Li:2024iyj,Li:2025fom}. Within the SM, these coefficients are known through the next-to-next-to-leading order in perturbative QCD~\cite{Gauld:2017tww}.

The examples quoted above focus on the lower moments $A_{0-4}$. The higher moments $A_{5-7}$ are less studied. They are much smaller in the SM, as they require both non-zero transverse momentum of the lepton pair and complex phases in the amplitude that first occur at loop level~\cite{Frederix:2014cba}. They are typically denoted as ``naive" $T$-odd coefficients due to their transformation under the discrete symmetries of the SM, a topic we discuss later in this manuscript. They are generated at ${\cal O}(\alpha_s{}^2)$ in QCD. A detailed SM calculation of these $T$-odd angular coefficients, including both $W^\pm$ and $\gamma/Z$ exchange, is given in~\cite{Lyubovitskij:2024civ}. Lower-energy measurements of these terms probe the transverse spin structure of the nucleon~\cite{Gamberg:2005ip}. The measurement of the complete set of the $A_i$ moments at higher energies has been performed at the LHC for both neutral and charged current lepton pair production~\cite{CMS:2015cyj,ATLAS:2016rnf,ATLAS:2025mlt}. As these coefficients can already be probed with the existing LHC data, the potential for improvement at the high-luminosity LHC (HL-LHC) is significant.

Our goal in this paper is to investigate the new physics potential of future HL-LHC measurements of the $A_{5-7}$ moments. We use the Standard Model Effective Field Theory (SMEFT) to parametrize the effects of heavy new physics. Previous work by one of us on the potential of the HL-LHC to probe SMEFT in the Drell-Yan process focused on the inclusive transverse momentum and invariant mass distributions~\cite{Boughezal:2022nof}. It was shown that these measurements are particularly sensitive to operators at the dimension-8 level in the SMEFT. This occurs because dimension-6 corrections to the transverse momentum distribution in the Drell-Yan process are generated by gluon emission from an external leg rather than directly from an interaction vertex, just like in the SM, and are therefore dominated by soft gluon emission. Emission directly from a contact interaction first occurs at the dimension-8 level. This work also showed that probes of dimension-8 effects through these measurements can serve as a diagnostic tool to help distinguish different UV completions of the SMEFT that lead to equivalent results at the dimension-6 level. We extend this analysis to study the naive $T$-odd coefficients at high transverse momentum and invariant mass. The first contribution to these observables arises from $CP$-odd dimension-8 operators in the SMEFT, allowing future DY measurements at the HL-LHC to probe a larger class of possible new physics effects. We show that a joint fit of the $A_6$ and $A_7$ moments at a future HL-LHC can probe new physics in the previously unexplored direction up to the 1-2 TeV level for certain Wilson coefficients in a fully-marginalized fit.

Our paper is organized as follows. We review these angular moments and discuss how discrete symmetries dictate their behavior in Sec.~\ref{sec:CSmoments}. We review the SMEFT in Sec.~\ref{sec:SMEFT}. Our focus is on the $CP$-odd operators that contribute to the naive $T$-odd moments $A_{6-7}$. We present our numerical results, including fits to simulated HL-LHC data, in Sec.~\ref{sec:numerics}. Finally, we conclude in Sec.~\ref{sec:conc}.

\section{Collins-Soper moments in Drell-Yan}
\label{sec:CSmoments}

\begin{figure}
    [H]
    \centering
    \includegraphics[width=.5\linewidth]{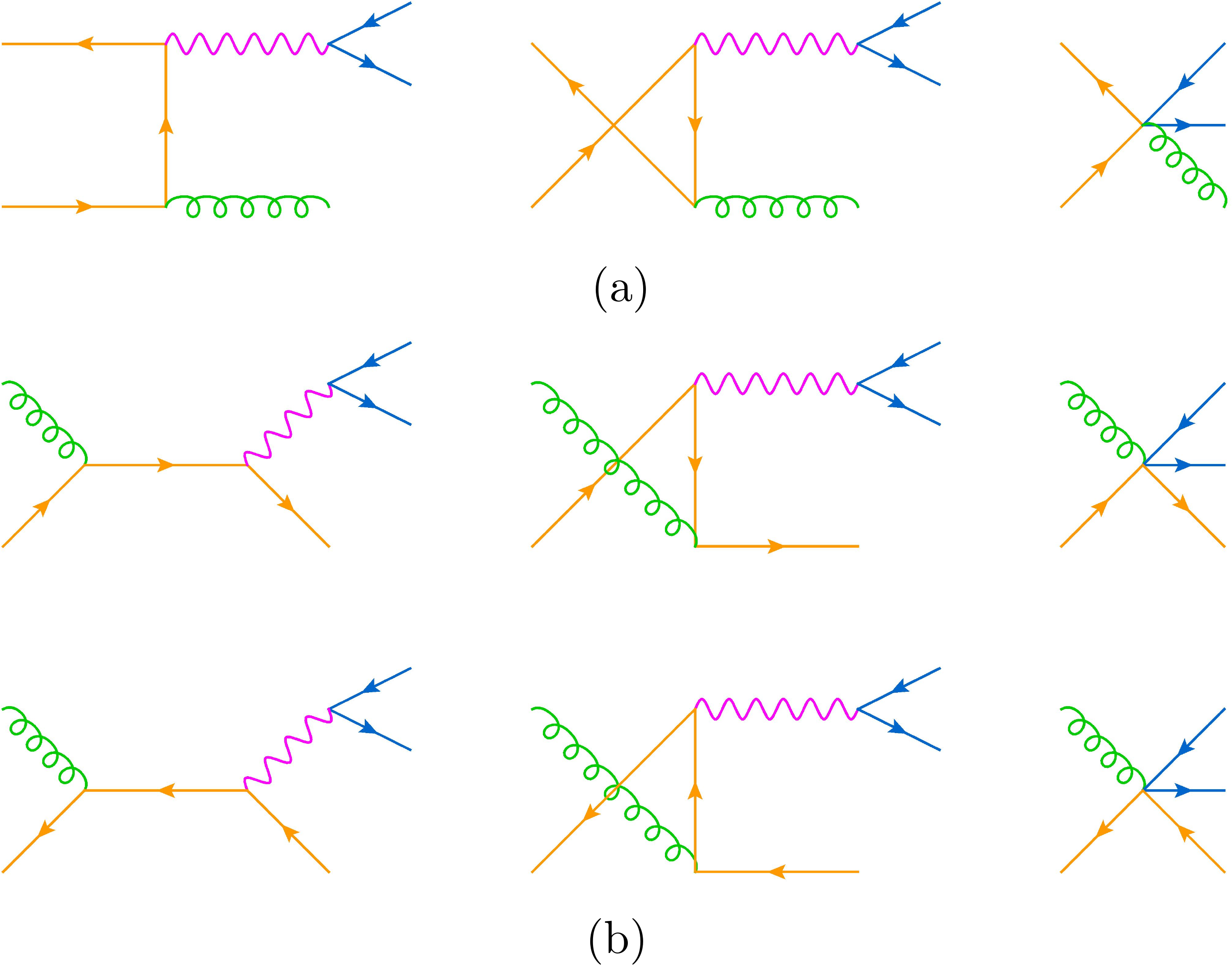}
    \caption{The tree-level Feynman diagrams describing the underlying partonic processes: (a) pair annihilation, (b) Compton scattering.}
    \label{fig:diagrams}
\end{figure}

We review here the theoretical formalism, including the calculation of the Drell-Yan (DY) process at finite transverse momentum and the relevant operators in the Standard Model Effective Field Theory (SMEFT). Production of a lepton pair at finite transverse momentum occurs through the hadronic process $pp \to j V \to je^-e^+$, where $V$ denotes either the photon or $Z$-boson. The underlying partonic processes at leading order in perturbative QCD are pair annihilation,
\begin{gather}
    q_i (p_a) + \bar q_i (p_b) \to g (p_1) + V(p_{23}) \to g (p_1) + e^- (p_2) + e^+ (p_3),
\end{gather}
and Compton scattering,
\begin{gather}
    q_i (p_a) + g (p_b) \to q_i (p_1) + V(p_{23}) \to q_i (p_1) + e^- (p_2) + e^+ (p_3), \\
    \bar q_j (p_a) + g (p_b) \to \bar q_j (p_1) + V(p_{23}) \to \bar q_j (p_1) + + e^- (p_2) + e^+ (p_3).
\end{gather}
For completeness, we display the tree-level Feynman diagrams in Fig.~\ref{fig:diagrams}. The first two columns show the SM diagrams, while the last column shows the SMEFT diagrams. We use {\sc FeynArts}~\cite{Hahn:2000kx} and~{\sc FeynCalc} \cite{Shtabovenko:2023idz, Shtabovenko:2020gxv, Shtabovenko:2016sxi, Mertig:1990an} for our analytic calculations of the partonic processes. 

Our main interest in this work is the angular distribution of the final-state lepton. As noted in the introduction, the differential cross section can be parametrized in terms of the Collins-Soper (CS) moments as~\cite{Collins:1977iv, AlcarazMaestre:2020, Alioli:2020kez}
\begin{gather}
    {\d\sigma \over \d \Omega^\star} = {3\sigma \over 16\pi} \bb{1 + c_{\theta^\star}{}^2 + \sum_{m=0}^7 A_m Y_m (\Omega^\star)},
\end{gather}
where $\Omega^\star = (\theta^\star, \varphi^\star)$ are the CS angles. The $Y_m$ are  linear combinations of the spherical harmonics $Y_1^0$, $Y_1^1 \pm Y_1^{-1}$, $Y_2^0$, $Y_2^1 \pm Y_2^{-1}$, and $Y_2^2 \pm Y_2^{-2}$:
\begin{gather}
    Y_0 = \frac12 (1 - 3 c_{\theta^\star}{}^2), \quad
    Y_1 = s_{2\theta^\star} c_{\varphi^\star}, \quad 
    Y_2 = \frac12 s_{\theta^\star}{}^2 c_{2 \varphi^\star}, \quad 
    Y_3 = s_{\theta^\star} c_{\varphi^\star}, \quad 
    Y_4 = c_{\theta^\star}, \\
    Y_5 = s_{\theta^\star}{}^2 s_{2\varphi^\star}, \quad
    Y_6 = s_{2\theta^\star} s_{\varphi^\star}, \quad
    Y_7 = s_{\theta^\star} s_{\varphi^\star},
    \label{eq:YLMdefs}
\end{gather}
with
\begin{gather}
    \int \d{\Omega^\star} \ Y_m Y_n \propto \delta_{mn}.
\end{gather}
The $A_m$ are the CS moments. $c$ and $s$ indicate the cosine and sine of the angle given in the subscript, respectively. Using the orthogonality of the spherical harmonics, we can obtain the CS moments separately as weighted averages over phase space with appropriate weight functions:
\begin{gather}
    A_0 = {20 \over 3} \langle Y_0 \rangle + \frac23, \quad
    A_1 = 5 \langle Y_1\rangle, \quad 
    A_2 = 20 \langle Y_2\rangle, \quad 
    A_3 = 4 \langle Y_3\rangle, \quad 
    A_4 = 4 \langle Y_4\rangle, \nonumber \\
    A_5 = 5 \langle Y_5\rangle, \quad 
    A_6 = 5 \langle Y_6\rangle, \quad 
    A_7 = 4 \langle Y_7\rangle,
\end{gather}
with
\begin{gather}
    \langle Y_m\rangle = {\int Y_m \d\sigma \over \sigma}.
\end{gather}
At non-zero transverse momenta the lepton angles are defined in the CS frame~\cite{Collins:1977iv}, in which the lepton pair is at rest.

The linear combinations of spherical harmonics defined in Eq.~\ref{eq:YLMdefs} have definite properties under discrete symmetries, which is useful in understanding their behavior in the Standard Model and beyond. Writing a raising operator for a particle with spin $s$ and 3-momentum $\vec{p}$ as $a^{s \dagger}_{\vec{p}}$, we can define the transformation of a state under parity and ``naive" time-reversal~\cite{Hagiwara:1982cq}:
\begin{eqnarray}
P (c \,a^{s \dagger}_{\vec{p}}) P^{-1} &=& c \,a^{s \dagger}_{-\vec{p}},\nonumber \\
A (c \,a^{s \dagger}_{\vec{p}}) A^{-1} &=& c \,a^{-s \dagger}_{-\vec{p}},
\end{eqnarray}
where $c$ is a $c$-number. The naive time-reversal operator $A$ differs from the normal time-reversal operator in that it is unitary, and does not conjugate the $c$-number. The $A_{5-7}$ structures in the angular distribution are odd under the combined $AP$ transformation, while the other terms are even. In QCD, which is $CP$-conserving, the only way to generate an $AP$-odd effect is through an imaginary part of a loop amplitude, which differentiates $AP$ from $CP$. Generating such a term at non-zero $p_T$ occurs first at ${\cal O}(\alpha_s{}^2)$ in QCD. In the SMEFT, there are dimension-8 $CP$-odd operators that contribute to the DY transverse momentum distribution preferentially at high-$p_T$, and can therefore potentially give sizeable shifts of the $A_{5-7}$ moments. These are the focus of our study.

\section{Review of the SMEFT}
\label{sec:SMEFT}

We review here the Standard Model Effective Field Theory (SMEFT), with a focus on the aspects needed for our study. The SMEFT is a model-independent extension of the Standard Model (SM) with operators $O_k^{(n)}$ of mass dimension $n>4$ suppressed by an ultraviolet scale $\Lambda$ at which new particles are expected to appear. The SMEFT uses the existing SM spectrum and maintains all SM gauge symmetries. The operators in the SMEFT are multiplied by Wilson coefficients $C_k^{(n)}$ that act as effective couplings. The SMEFT Lagrangian is schematically given by 
\begin{gather} 
    \mathcal L = \mathcal L_{\rm SM} + \sum_{n>4} {1 \over \Lambda^{n-4}} \sum_k C_k^{(n)} O_k^{(n)}. 
\end{gather} 
In our work, we focus on the case $n = 8$ since it is at this order that $CP$-odd operators contributing to the $A_{5-7}$ Collins-Soper (CS) moments first appear. See~\cite{Brivio:2017vri} for a review of the SMEFT. 

As discussed previously, we are interested in SMEFT operators that satisfy two criteria: they lead to $CP$-odd contributions to Drell-Yan (as we study tree-level results here, there are no significant $AP$-odd effects that are also $CP$-even), and they generate a transverse momentum for the lepton pair directly from a hard vertex rather than from emission off an external leg. These two criteria can potentially lead to sizable effects for the $A_{5-7}$ CS moments. These necessarily come from four-fermion operators with an explicit gluon field-strength tensor, as discussed in Ref.~\cite{Boughezal:2022nof}, which we briefly review here. The standard discrete transformation rules for the various fermion bilinears and other structures that appear in the operators can be written as
\begin{gather}
CP (\bar \psi \gamma^\mu \psi) (CP)^{-1} = -(-1)^\mu (\bar \psi \gamma^\mu \psi), \\
CP (\partial_\mu) (CP)^{-1} = (-1)^\mu \partial_\mu, \\
CP(A_\mu) (CP)^{-1} = -(-1)^\mu A_\mu. 
\end{gather}
The gauge field contais an additional minus sign due to its charge conjugation properties. We recall that $(-1)^\mu = 1$ for $\mu = 0$ and $-1$ otherwise, according to the definition in~\cite{Peskin:1995ev}. We can apply this to the operators that appear to get the following transformations:
\begin{gather}
CP \{(\bar \ell \gamma^\mu \ell) (\bar q \gamma^\nu q) G_{\mu\nu}\} (CP)^{-1} = -[(-1)^\mu]^2 [(-1)^\nu]^2 (\bar \ell \gamma^\mu \ell) (\bar q \gamma^\nu q) G_{\mu\nu}, \label{ex1} \\ 
CP \{(\bar \ell \gamma^\mu \ell) (\bar q \gamma^\nu q)  \epsilon_{\mu\nu\rho\sigma} G^{\rho\sigma}\} (CP)^{-1} = -[(-1)^\mu (-1)^\nu (-1)^\rho (-1)^\sigma] (\bar \ell \gamma^\mu \ell) (\bar q \gamma^\nu q) \epsilon_{\mu\nu\rho\sigma} G^{\rho\sigma}, \label{ex2}
\end{gather}
where we have chosen a representative operator from 
the dimension-8 SMEFT operators of interest presented in Table~\ref{tab:smeft_ops}~\cite{Murphy:2020rsh, Li:2020gnx}. In \eqref{ex1}, the term in the square brackets is unity, since both minus sign factors appear squared. In \eqref{ex2}, we note that three of the indices $\mu$, $\nu$, $\rho$, and $\sigma$ will be spatial due to the Levi-Civita tensor, and thus the factor in the square bracket is negative for this operator. We obtain
\begin{gather}
CP [(\bar \ell \gamma^\mu \ell) (\bar q \gamma^\nu q) G_{\mu\nu}] (CP)^{-1} = - (\bar \ell \gamma^\mu \ell) (\bar q \gamma^\nu q) G_{\mu\nu}, \\ 
CP \{(\bar \ell \gamma^\mu \ell) (\bar q \gamma^\nu q)  \epsilon_{\mu\nu\rho\sigma} G^{\rho\sigma}\} (CP)^{-1} = + (\bar \ell \gamma^\mu \ell) (\bar q \gamma^\nu q)  \epsilon_{\mu\nu\rho\sigma} G^{\rho\sigma}.
\end{gather}
We also note that we have previously checked in Ref.~\cite{Boughezal:2022nof} that matching a $CP$-even UV theory, in the example of a model with a vector leptoquark studied there, leads to the $CP$-even operators. The explicit calculation is presented there. 
\begin{table}
    [H]
    \centering
    \caption{Dimension-8 four-fermion operators with a gluon field-strength tensor that contribute to the Drell-Yan transverse momentum spectrum classified according to their $CP$ signature.}
    \label{tab:smeft_ops}
    {\renewcommand{\arraystretch}{2}%
    \begin{tabular}{|c|c|c|c|}
        \hline
        \multicolumn{2}{|c|}{$CP$ even} & \multicolumn{2}{c|}{$CP$ odd} \\
        \hline 
        $\tilde O_{\ell^2 q^2 g}^{(1)}$ & $(\bar \ell \gamma^\mu \ell) (\bar q \gamma^\nu T^A q) \tilde G^A_{\mu\nu}$ & $O_{\ell^2 q^2 g}^{(1)}$ & $(\bar \ell \gamma^\mu \ell) (\bar q \gamma^\nu T^A q) G^A_{\mu\nu}$ \\
        $\tilde O_{\ell^2 q^2 g}^{(3)}$ & $(\bar \ell \tau^i \gamma^\mu \ell) (\bar q \tau^i \gamma^\nu T^A q) \tilde G^A_{\mu\nu}$ & $O_{\ell^2 q^2 g}^{(3)}$ & $(\bar \ell \tau^i \gamma^\mu \ell) (\bar q \tau^i \gamma^\nu T^A q) G^A_{\mu\nu}$ \\
        $\tilde O_{e^2 u^2 g}$ & $(\bar e \gamma^\mu e) (\bar u \gamma^\nu T^A u) \tilde G^A_{\mu\nu}$ & $O_{e^2 u^2 g}$ & $(\bar e \gamma^\mu e) (\bar u \gamma^\nu T^A u) G^A_{\mu\nu}$ \\
        $\tilde O_{e^2 d^2 g}$ & $(\bar e \gamma^\mu e) (\bar d \gamma^\nu T^A d) \tilde G^A_{\mu\nu}$ & $O_{e^2 d^2 g}$ & $(\bar e \gamma^\mu e) (\bar d \gamma^\nu T^A d) G^A_{\mu\nu}$ \\
        $\tilde O_{\ell^2 u^2 g}$ & $(\bar \ell \gamma^\mu \ell) (\bar u \gamma^\nu T^A u) \tilde G^A_{\mu\nu}$ & $O_{\ell^2 u^2 g}$ & $(\bar \ell \gamma^\mu \ell) (\bar u \gamma^\nu T^A u) G^A_{\mu\nu}$ \\
        $\tilde O_{\ell^2 d^2 g}$ & $(\bar \ell \gamma^\mu \ell) (\bar d \gamma^\nu T^A d) \tilde G^A_{\mu\nu}$ & $O_{\ell^2 d^2 g}$ & $(\bar \ell \gamma^\mu \ell) (\bar d \gamma^\nu T^A d) G^A_{\mu\nu}$ \\
        $\tilde O_{q^2 e^2 g}$ & $(\bar e \gamma^\mu e) (\bar q \gamma^\nu T^A q) \tilde G^A_{\mu\nu}$ & $O_{q^2 e^2 g}$ & $(\bar e \gamma^\mu e) (\bar q \gamma^\nu T^A q) G^A_{\mu\nu}$ \\
        \hline 
    \end{tabular}}
\end{table}

In this table, $\ell$ and $q$ are left-handed SU(2) doublets and $e$, $u$, and $d$ are right-handed singlets; in what follows, we switch back to the usual notation that says $\ell$ ($q$) is a Dirac lepton (quark) that appears in the partonic processes. The $CP$-even operators affect high transverse momentum DY production but do not generate the CS moments of interest. The potential for the high-luminosity LHC (HL-LHC) to probe the $CP$-even operators, and how they can help distinguish between different ultraviolet completions of the SMEFT, was discussed in~\cite{Boughezal:2022nof}. 

Before presenting numerical results, we make a few comments about these operators.
\begin{itemize}

    \item We do not obtain novel angular structures requiring spherical harmonics with $l >2$ from these operators. Such effects occur only for certain dimension-8 four-fermion operators that have non-zero $l=2$ partial waves at the amplitude level and therefore contribute to $l=3$ partial waves in the cross section, as shown in~\cite{Alioli:2020kez}.

    \item The Lam-Tung relation~\cite{Lam:1978pu} $A_0 = A_2$ is satisfied at tree-level in the presence of these operators.

    \item The moments $A_5$, $A_6$, and $A_7$ are nonzero only starting at ${\cal O}(\alpha_s{}^2)$ in the SM. They are generated at tree-level by the $CP$-odd category of operators.
    
    \item We find that SMEFT contributions to $A_5$ are negligible compared to $A_6$ and $A_7$. This suppression arises from the partonic channel structure: for the $q\bar{q}$ initial state, $A_5$ vanishes identically due to exact antisymmetry under the two scattering angle solutions, while $gq$ initial states yield nonzero but small contributions, suppressed by two to three orders of magnitude relative to $A_6$ and $A_7$. We therefore do not consider $A_5$ further in our study.
    
\end{itemize}
The SM results for the $A_6$ and $A_7$ moments, together with a comparison to the experimental data, are shown in Ref.~\cite{ATLAS:2016rnf}. The coefficients are small in the SM, as expected since they begin only at ${\cal O}(\alpha_s{}^2)$. We focus here on using $A_6$ and $A_7$ to probe potential $CP$-odd heavy new physics effects. 

\section{Numerical results}
\label{sec:numerics}

We present here our numerical results, beginning by examining the size of each operator's contribution to various Drell-Yan (DY) observables. We then present fits to simulated high-luminosity LHC (HL-LHC) data. We first discuss our conventions. We define the following vertex factors which enter the Standard Model (SM) couplings:
\begin{gather}
    C_{ff\gamma} = -e Q_f, \quad
    C_{ffZ}^{L/R} = g_Z g_{L/Rf}, \quad
    C_{qqg} = -g_s,
\end{gather}
where
\begin{gather}
    g_Z = {e \over s_W c_W}, \quad  
    g_{Lf} = {T_{3f} - s_W{}^2 Q_f}, \quad
    g_{Rf} = {- s_W{}^2 Q_f}, \quad 
    g_s = \sqrt{4 \pi \alpha_s}.
\end{gather}
We employ an electroweak input scheme with $\{G_F, s_W{}^2, m_Z\}$ as input parameters. We have
\begin{gather}
    G_F = 1.1663787 \times 10^{-5} \ \GeV^{-2}, \quad
    s_W{}^2 = 0.23113, \quad
    m_Z = 91.1876 \ \GeV. 
\end{gather}
The remaining parameters are derived from these assuming tree-level SM relations. Our unusual choice of $s_W^2$ as an input parameter incorporates radiative corrections into the effective $Z$-boson vertices that are important when considering parity-violating effects such as those considered here. We assume $N_f=5$ and use the the NNPDF3.1 NLO PDF set~\cite{NNPDF:2017mvq} obtained through {\sc LHAPDF}~\cite{Buckley:2014ana}. We validate our SM results with {\sc MadGraph}~\cite{Alwall:2014hca} and~{\sc MCFM}~\cite{Campbell:2019dru}.

\begin{figure}
[H]
\centering
\includegraphics[width=\textwidth]{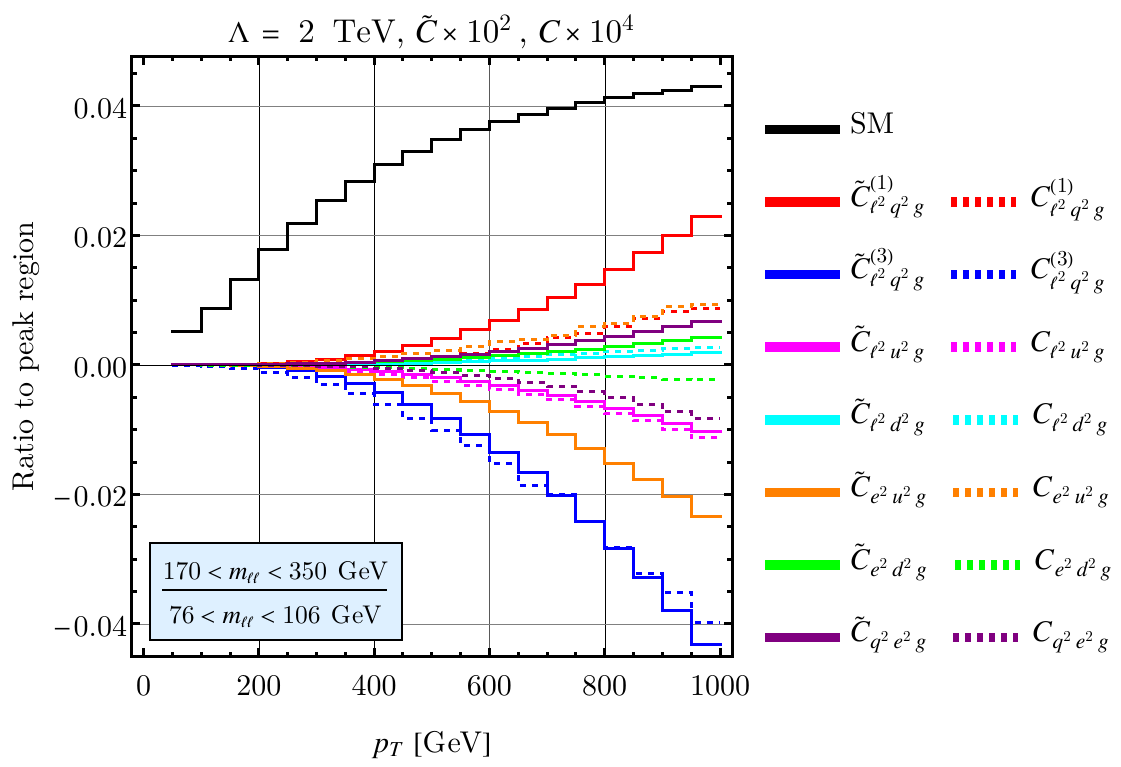}
\caption{The ratios of the inclusive SM and SMEFT cross sections to the $Z$-peak for the invariant mass bin $m_{\ell\ell}=[170, 350]$~GeV. The effective scale is set $\Lambda = 2 \ \TeV$, and the SMEFT contributions are amplified by a factor of 100 for the $CP$-even operators and 10,000 for the $CP$-odd operators. The nonzero $CP$-odd interference is a finite $Z$ width effect and vanishes for $\Gamma_Z\to0$.}
\label{fig:smeft_contributions_to_xs_1}
\end{figure}

\begin{figure}
[H]
\centering
\includegraphics[width=\textwidth]{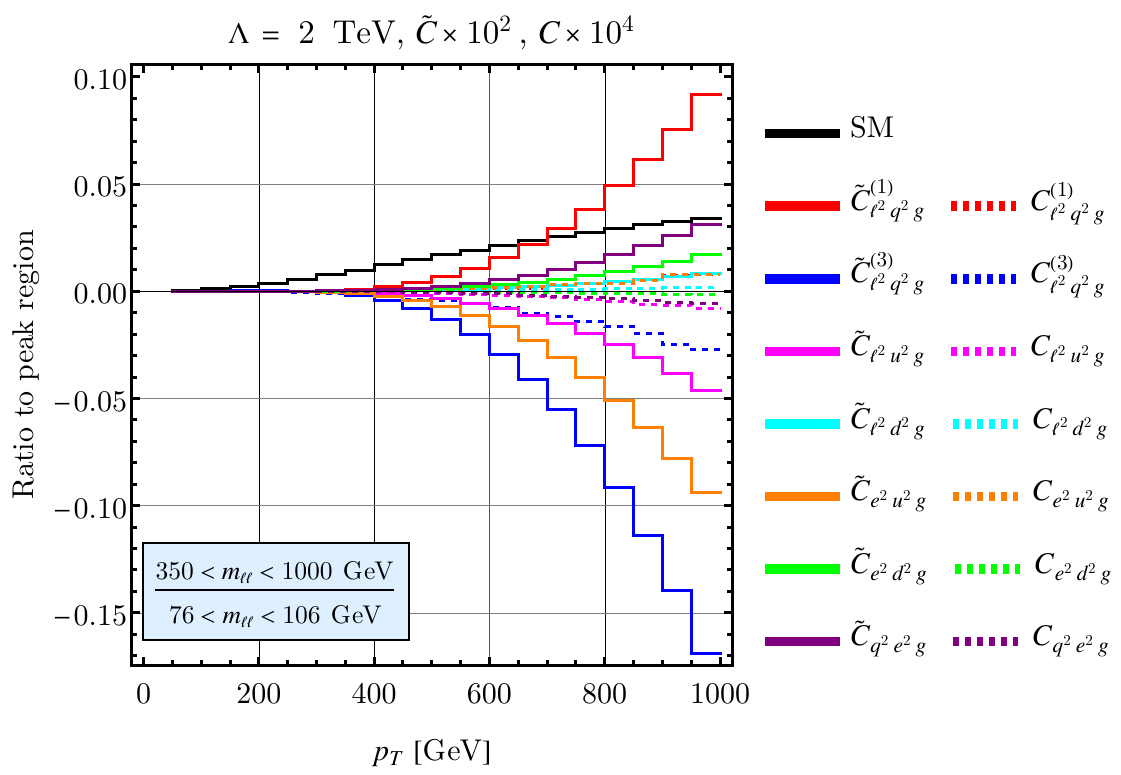}
\caption{The same as Fig.~\ref{fig:smeft_contributions_to_xs_1} but for $m_{\ell\ell} = [350, 1000]$ GeV.}
\label{fig:smeft_contributions_to_xs_2}
\end{figure}

To begin our study of the SMEFT contributions to the DY cross section, we write our hadronic cross section as
\begin{gather}
    \sigma = \sigma_{\rm SM} + \sum_i C_i \sigma_i + \sum_i \widetilde C_i \widetilde \sigma_i.
\end{gather}
We define the SM cross section integrated around the $Z$-peak  $76 < \mll < 106 \ {\rm GeV}$ as $\sigma_{\rm SM}^Z$. We activate the Wilson coefficients one at a time by setting $C_i = 1$ or $\widetilde C_i = 1$, and plot the SMEFT shifts of the cross section for invariant mass bins beyond the $Z$-peak as a function of $\pt$, normalized to the $Z$-peak cross section. We call this the ratio to the peak and denote it by $\sigma_k/\sigma_{\rm SM}^Z$, where $\sigma_k$ is any one of $\sigma_{\rm SM}$, $\sigma_i$, and $\widetilde \sigma_i$. In Figs.~\ref{fig:smeft_contributions_to_xs_1} and~\ref{fig:smeft_contributions_to_xs_2}, we plot the ratios to the peak for the invariant mass bins [170, 350] GeV and [350, 1000] GeV in conjunction with the $\pt$ bins [50, 100], [100, 150], \ldots, [950, 1000] GeV. We note that these are inclusive observables with the lepton angles fully integrated over. In the limit $\Gamma_Z \to 0$, the SM-SMEFT interference terms from $CP$-odd operators vanish in these inclusive cross sections. This cancellation occurs because the observables are only naively $T$-odd: without an absorptive contribution in the amplitude, the interference integrates to zero. In our calculation, we retain the finite $Z$ width through the propagator $1/(q^2 - m_Z^2 + im_Z\Gamma_Z)$, which provides the necessary absorptive part and prevents exact cancellation. The integrated SM-SMEFT interference from $CP$-odd operators therefore becomes nonzero but width-suppressed. We have verified explicitly that setting $\Gamma_Z \to 0$ causes these linear $CP$-odd contributions to vanish, with the cancellation occurring already upon integration over the Collins-Soper azimuthal angle $\varphi$ alone. The nonzero values visible in Figs.~\ref{fig:smeft_contributions_to_xs_1} and \ref{fig:smeft_contributions_to_xs_2} therefore originate entirely from the finite width and are numerically small. In these figures, the black lines are the SM terms and the colored lines are the SMEFT corrections. The solid color lines are the Wilson coefficients corresponding to the $CP$-even operators and the dashed ones are those corresponding to the $CP$-odd operators. We assume a center-of-mass energy of $\sqrt s = 13 \ {\rm TeV}$, we set $\Lambda = 2 \ {\rm TeV}$, and we enhance the SMEFT contributions $\widetilde \sigma_i$ by a factor of $10^2$ and $\sigma_i$ by $10^4$ for display purposes. 


We can observe the following facts from these plots.
\begin{itemize}
    \item The SMEFT corrections become sizable for higher $\pt$ values. 
    \item The SMEFT corrections characterized by Wilson coefficients corresponding to the $CP$-even operators are significantly larger compared to the $CP$-odd operators. The $CP$-odd operators become important only when a $CP$-odd observable is measured.
    \item The SMEFT corrections characterized by the operators involving the interaction of two left-handed currents lead to the largest shifts from the SM results. 
\end{itemize}
We note that the $CP$-odd operators leads to corrections too small to see in these observables; other measurements are needed to observe their effects.

\begin{figure}
[H]
\centering
\includegraphics[width=.4875\textwidth]{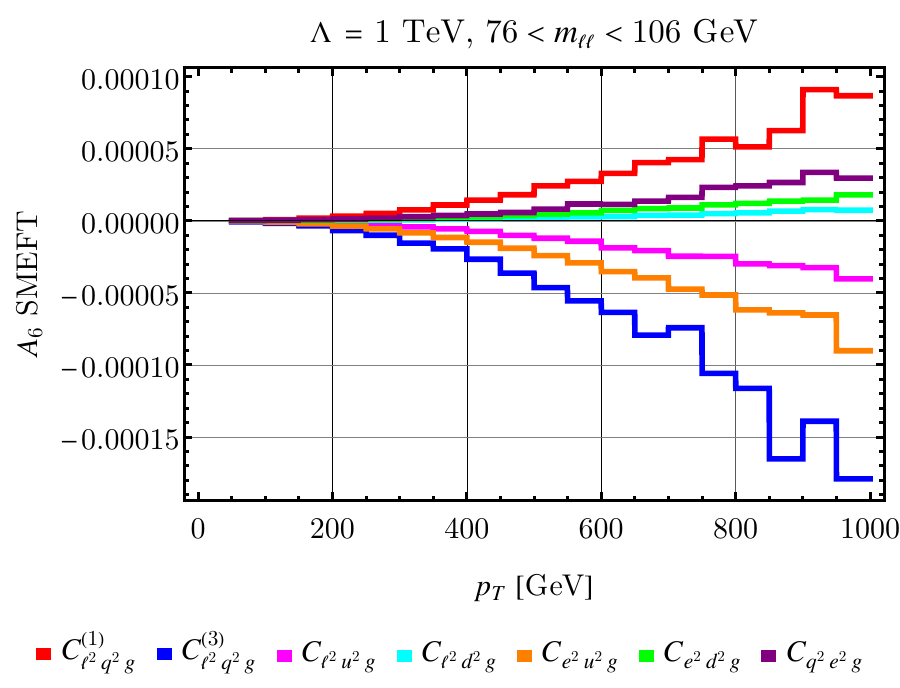}
\includegraphics[width=.4875\textwidth]{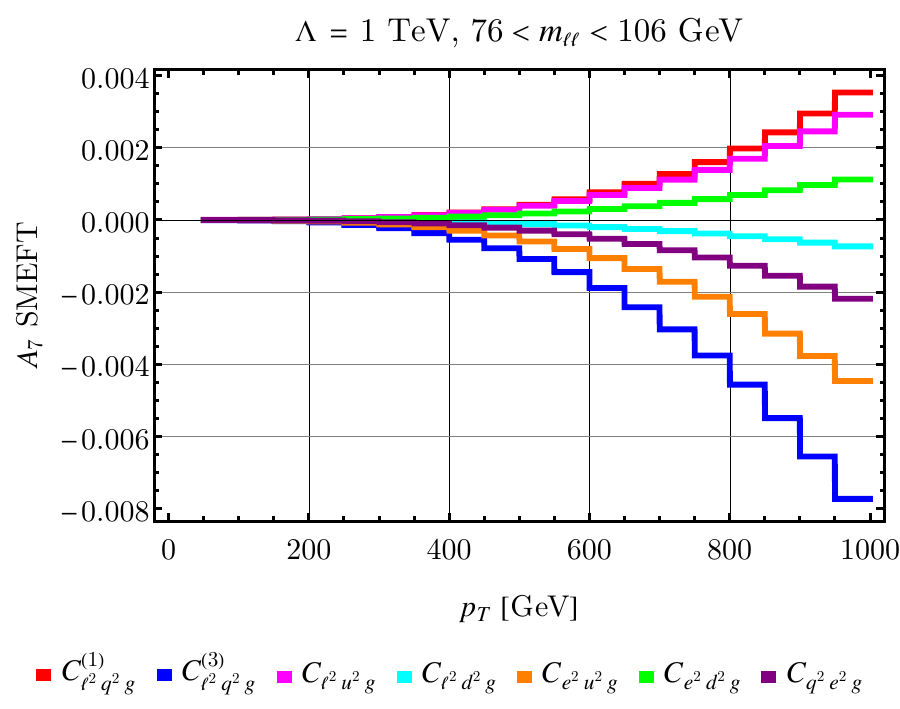}
\caption{SMEFT corrections to the $A_6$ and $A_7$ moments as a function of $p_T$ in the invariant mass bin $76 < m_{ll} < 106$ GeV.}
\label{fig:bin1}
\end{figure}

\begin{figure}
[H]
\centering
\includegraphics[width=.4875\textwidth]{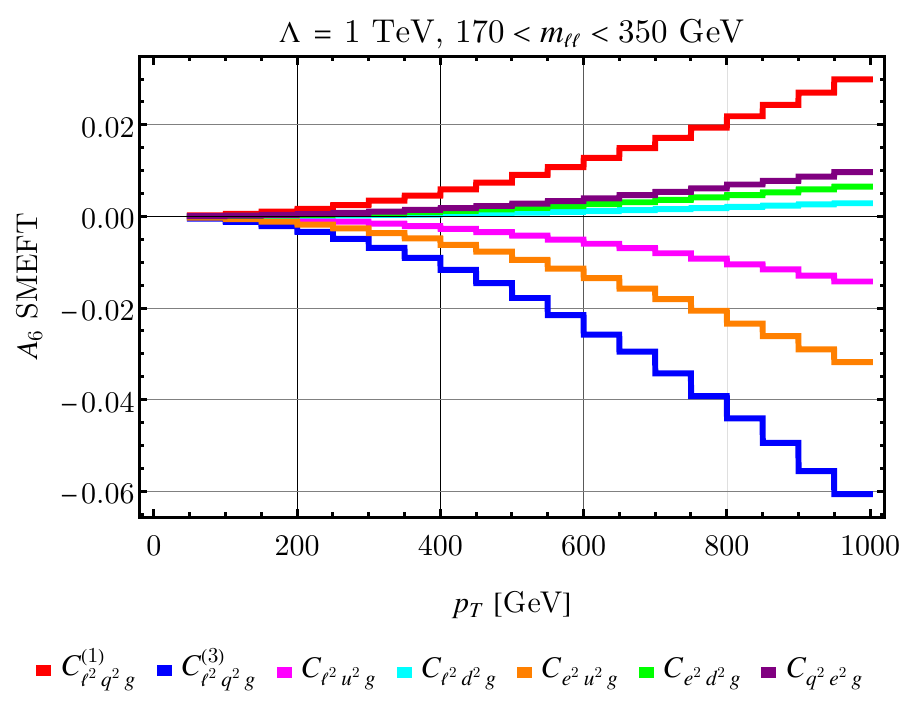}
\includegraphics[width=.4875\textwidth]{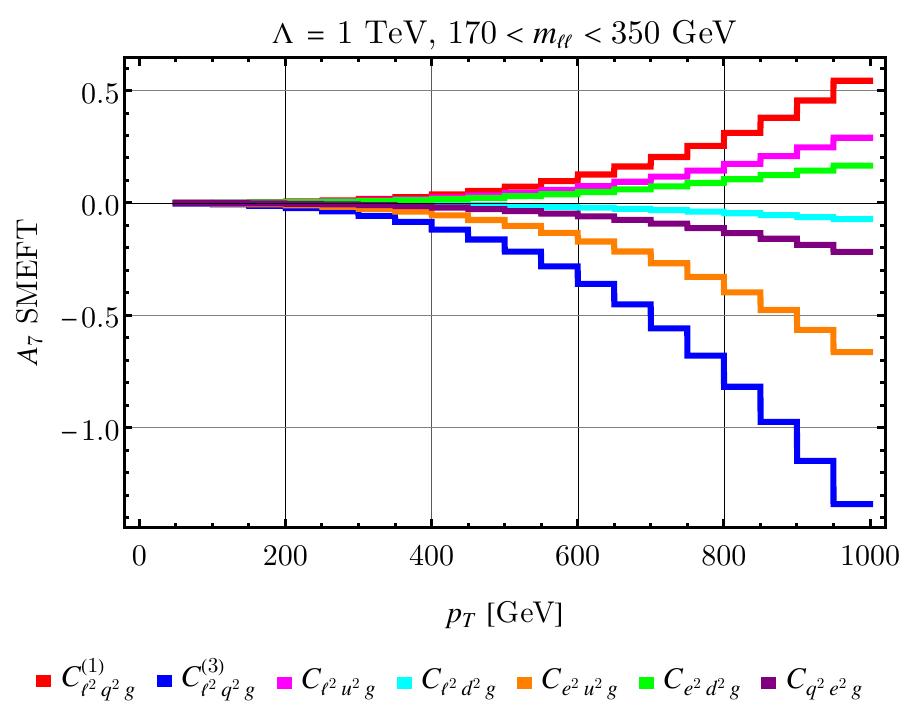}
\caption{The same as Fig.~\ref{fig:bin1} but for $170 < m_{\ell\ell} < 350$ GeV.}
\label{fig:bin2}
\end{figure}

\begin{figure}
[H]
\centering
\includegraphics[width=.4875\textwidth]{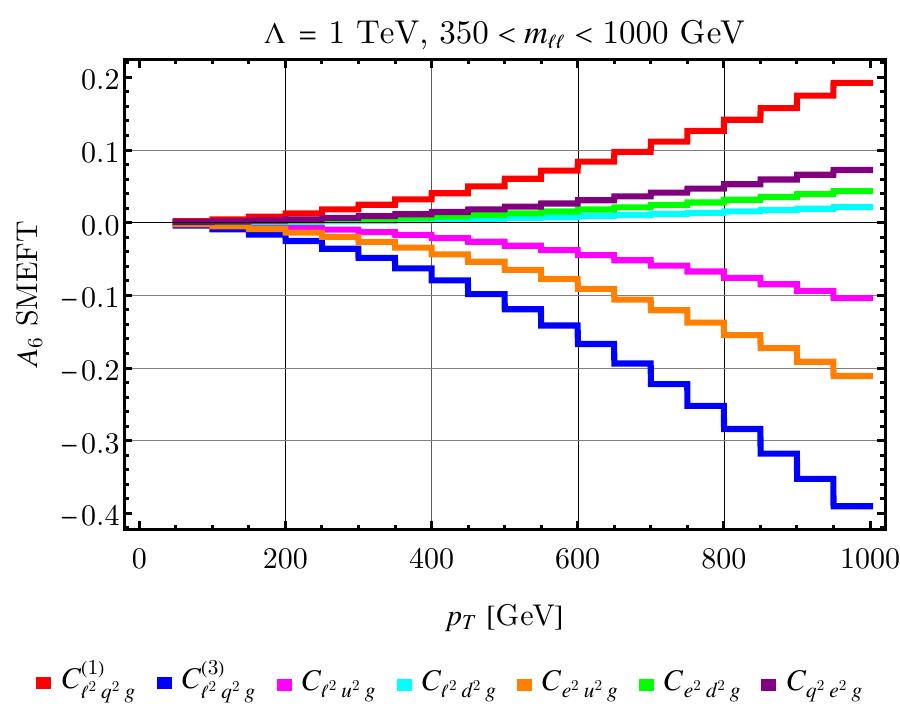}
\includegraphics[width=.4875\textwidth]{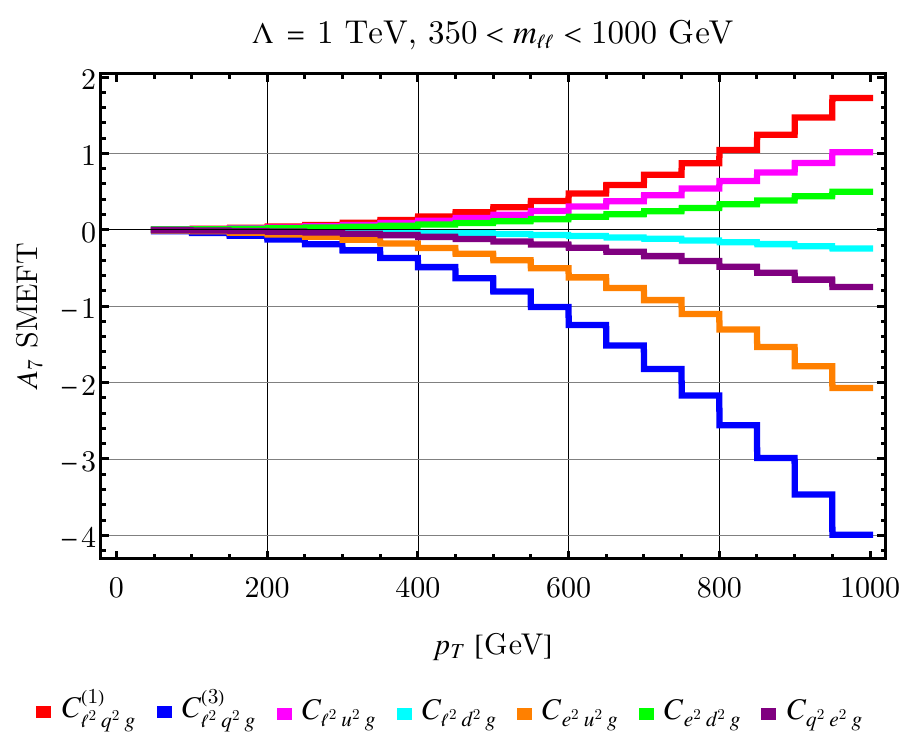}
\caption{The same as Fig.~\ref{fig:bin1} but for $350 < m_{\ell\ell} < 1000$ GeV.}
\label{fig:bin3}
\end{figure}

We next investigate the SMEFT corrections to the Collins-Soper (CS) moments as functions of $\mll$ and $\pt$, which is our primary interest in this paper. We note that any given CS moment is essentially the ratio of two cross-section integrals, one with a particular angular structure and one with unit weight, both of which can be written as a SM part plus some correction linearly proportional to a Wilson coefficient:
\begin{gather}
    A = {\int Y \d\sigma \over \sigma} = {N^{(0)} + C N^{(1)} \over D^{(0)} + C D^{(1)}},
\end{gather}
where $A$ is a CS moment, $Y$ is the corresponding angular structure, and $C$ is some Wilson coefficient. We linearize this in SMEFT Wilson coefficients by writing
\begin{gather}
    A = A^{(0)} + C A^{(1)}.
\end{gather}
We present in Figs.~\ref{fig:bin1},~\ref{fig:bin2}, and~\ref{fig:bin3} plots demonstrating $A^{(1)}$ for $A_6$ and $A_7$ for three different choices of invariant mass bins The colors correspond to the Wilson coefficients indicated in the captions; the choices match our color choices for the inclusive observables presented previously. This time, we exclusively focus on the $C$-operators because the $\widetilde C$-operators vanish for these observables. We see that the corrections become large at high transverse momentum and mass. The left-handed Wilson coefficient $C_{lq}$ generically gives the largest contribution, but the deviations from all operators are large enough to be probed with future HL-LHC data.


\subsection{HL-LHC simulation}

In this section, we perform fits to simulated HL-LHC data to determine how well these $CP$-odd operators can be probed by future HL-LHC data. We set the collider energy to $\sqrt s = 14 \ \TeV$ and assume an integrated luminosity of $\mathcal L = 3 \ {\rm ab}^{-1}$, consistent with current expectations~\cite{Bruning:2025pmh}. We employ the following cuts to mimic a realistic experimental acceptance:
\begin{itemize}
    \item leading electron: $\pt > 25 \ \GeV$,
    \item subleading electron: $\pt > 20 \ \GeV$,
    \item both electrons: $|\eta| < 2.4$,
    \item dilepton system: $\pt > 100 \ \GeV$, $|y| < 2.4$.
\end{itemize}
We assume the binning defined in~Table~\ref{tab:fine_bins}, chosen so that the relative statistical uncertainties in each bin are smaller than 10\%. This choice of binning is consistent with previous studies~\cite{Boughezal:2022nof}. We assume only statistical errors in our fit, consistent with the dominance of statistical uncertainties in the existing ATLAS analysis of the higher CS moments~\cite{ATLAS:2016rnf}. Since the SM predictions for the $A_{6-7}$ moments are small, systematic errors defined relative to the value of these moments will be subdominant to the statistical errors.
\begin{table}
    [H]
    \centering 
    \caption{Binning in invariant mass and transverse momemntum used in our numerical study.}
    \label{tab:fine_bins}
    \begin{tabular}{|l|p{10cm}|}
        \hline 
        $\mll$ [GeV] & $\pt$ [GeV] \\ 
        \hline 
        [300, 360] & [100, 110, 120, 130, 140, 150, 160, 170, 180, 190, 200, 210, 220, 230, 250, 270, 290, 310, 330, 350, 370, 400, 420, 440, 470, 500, 530, 560, 600, 660, 760, 7000] \\
        \hline 
        [360, 450] & [100, 110, 120, 130, 140, 150, 160, 170, 180, 190, 200, 210, 220, 240, 260, 290, 310, 330, 350, 370, 390, 410, 440, 470, 500, 530, 560, 610, 670, 770, 7000] \\
        \hline 
        [450, 600] & [100, 110, 120, 130, 140, 150, 160, 190, 210, 230, 250, 270, 290, 320, 340, 370, 390, 420, 460, 490, 520, 550, 580, 620, 680, 780, 7000] \\ 
        \hline 
        [600, 800] & [100, 110, 120, 130, 150, 170, 200, 220, 240, 260, 280, 310, 340, 380, 410, 440, 470, 510, 550, 620, 730, 7000] \\ 
        \hline 
        [800, 1100] & [100, 110, 120, 140, 160, 180, 200, 220, 250, 270, 300, 330, 360, 410, 460, 540, 660, 7000] \\ 
        \hline 
        [1100, 1500] & [100, 130, 160, 190, 230, 270, 320, 400, 520, 7000] \\
        \hline 
        [1500, 2000] & [100, 210, 330, 7000] \\ 
        \hline
        [2000, 2600] & [100, 7000] \\
        \hline 
    \end{tabular}
\end{table}

\begin{figure}
    [htbp]
    \centering
    \includegraphics[width=.7\textwidth]{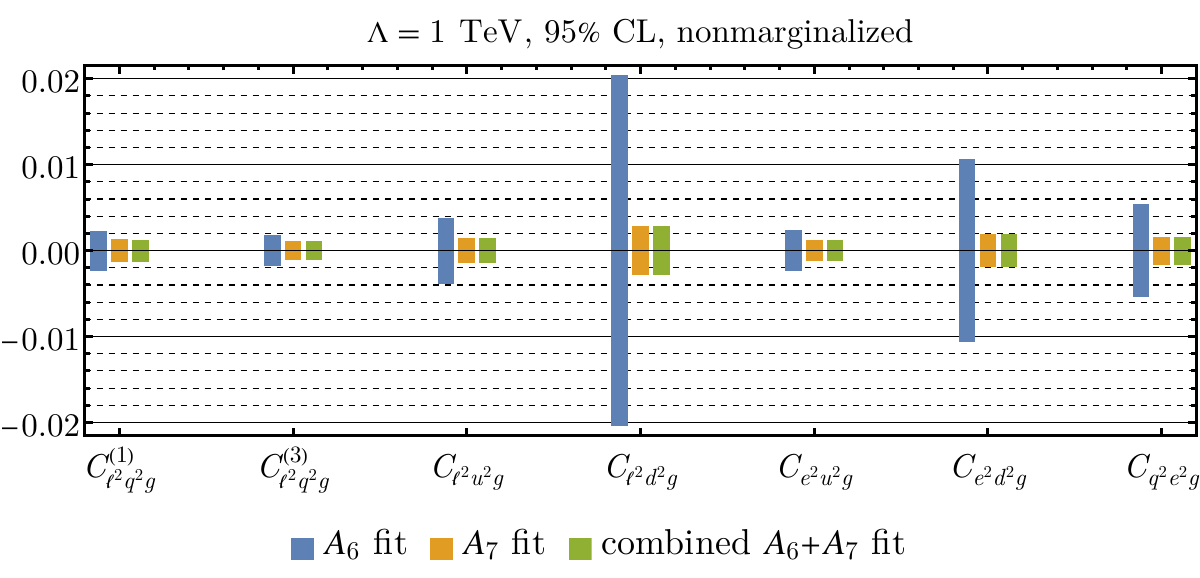}
    \caption{The 95\% CL bounds on the Wilson coefficients from a non-marginalized fit. }
    \label{fig:nm_bounds}
\end{figure}

\begin{figure}
    [htbp]
    \centering
    \includegraphics[width=.7\textwidth]{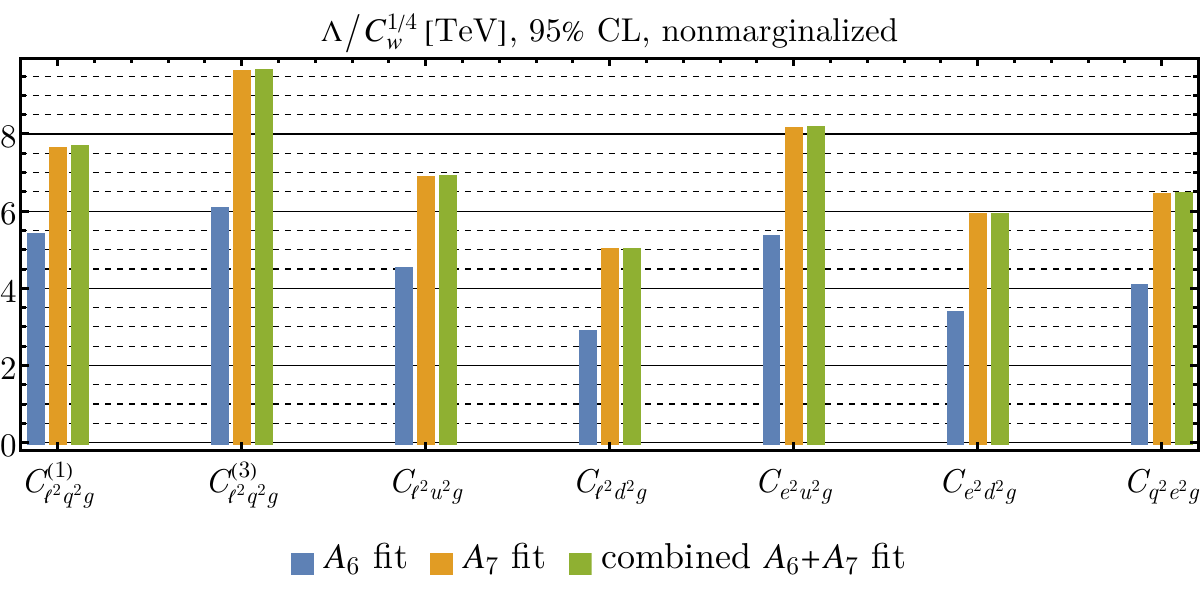}
    \caption{The 95\% CL bounds on the effective UV scales from a non-marginalized fit.}
    \label{fig:nm_effscales}
\end{figure}

\begin{figure}
    [htbp]
    \centering
    \includegraphics[width=.7\textwidth]{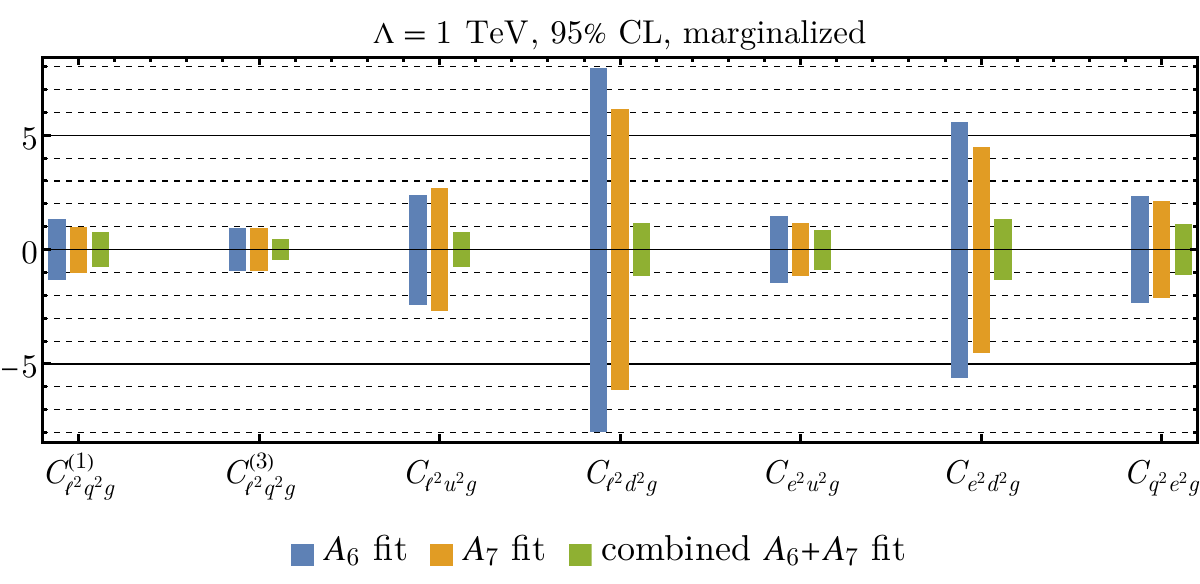}
    \caption{The 95\% CL bounds on the Wilson coefficients from a marginalized fit. }
    \label{fig:m_bounds}
\end{figure}

\begin{figure}
    [htbp]
    \centering
    \includegraphics[width=.7\textwidth]{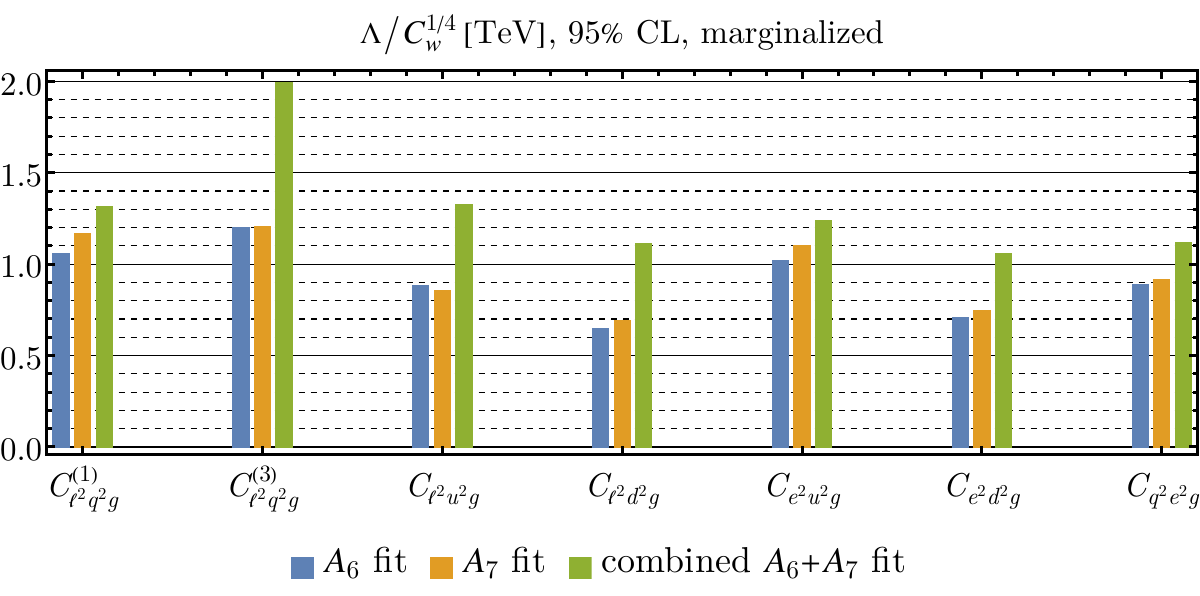}
    \caption{The 95\% CL bounds on the effective UV scales from a marginalized fit.}
    \label{fig:m_effscales}
\end{figure}

We present the non-marginalized bounds at 95\% confidence level (CL) with only a single Wilson coefficient activated in Fig.~\ref{fig:nm_bounds}. The corresponding effective ultraviolet (UV) scales, defined as the reparametrization invariant combinations $\Lambda/C_i^{1/4}$, are presented in Fig.~\ref{fig:nm_effscales}. In Fig.~\ref{fig:m_bounds}, we present the marginalized 95\% CL bounds and the corresponding effective UV scales in Fig.~\ref{fig:m_effscales}. From these figures, we note the following points.
\begin{itemize}
    \item The $A_6$ fits yield weaker bounds on the Wilson coefficients of interest compared to the $A_7$ fits. 
    
    \item The results of the combined fit resemble the $A_7$ fits.
    
    \item In $1d$ fits, we can probe effective scales up to 9 TeV.
    
    \item Once we activate all the Wilson coefficients (so as to perform a $7d$ fit), we see that the allowed intervals grow dramatically. This signals strong degeneracies between the SMEFT parameters.
    
\end{itemize}


\begin{figure}
    [H]
    \centering
    \includegraphics[width=.4875\textwidth]{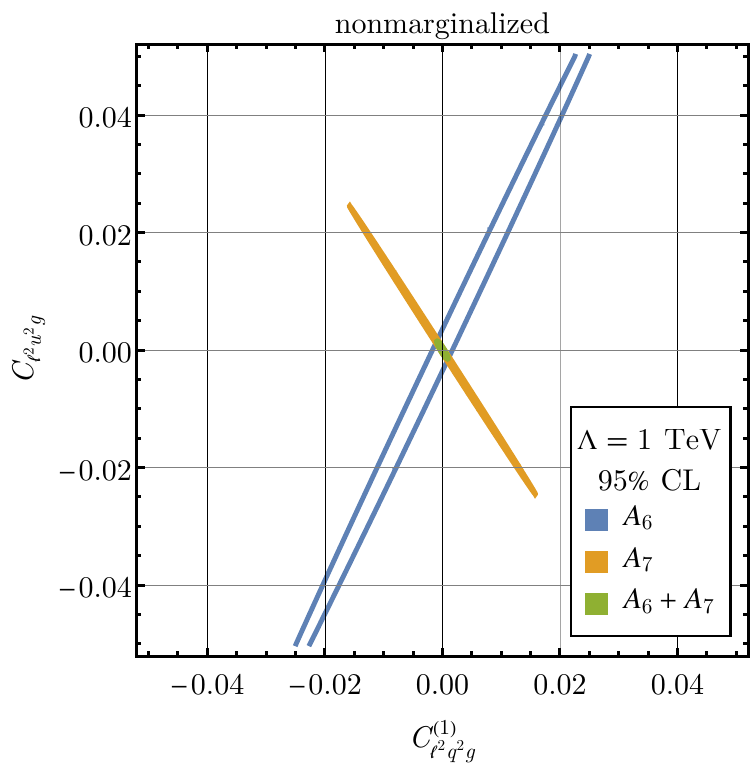}
    \includegraphics[width=.4875\textwidth]{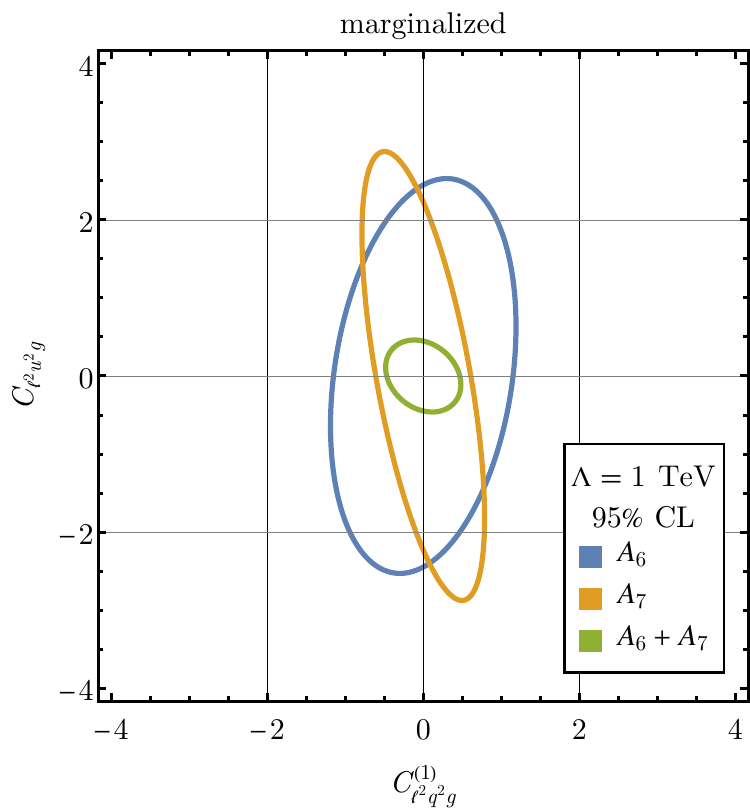}
    \caption{The 95\% confidence ellipses for both the non-marginalized and marginalized cases for the Wilson coefficients $C_{\ell^2 q^2 g}^{(1)}$ and $C_{\ell^2 u^2 g}$.}
    \label{fig:Wil2d}
\end{figure}

\begin{figure}
    [H]
    \centering
    \includegraphics[width=.4875\textwidth]{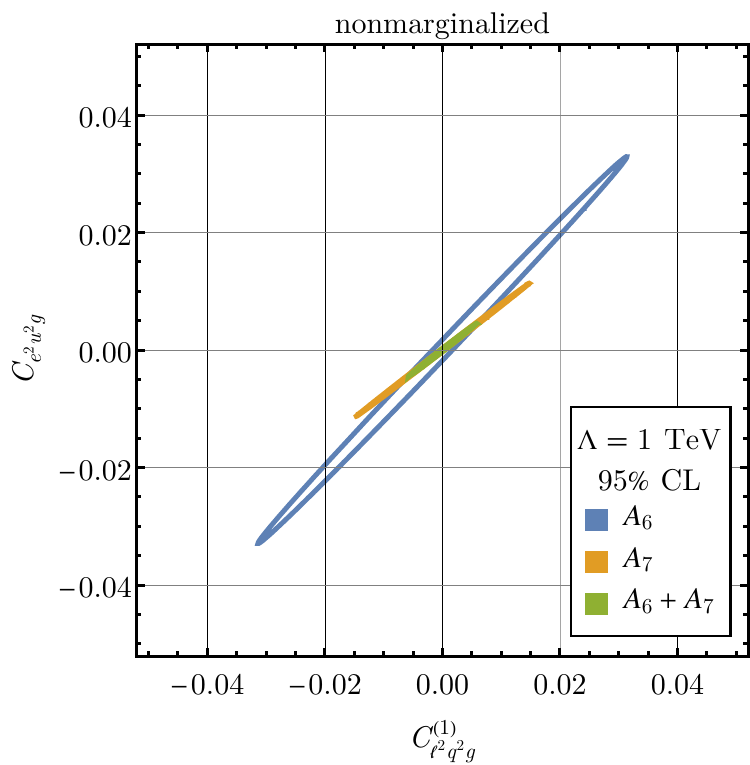}
    \includegraphics[width=.4875\textwidth]{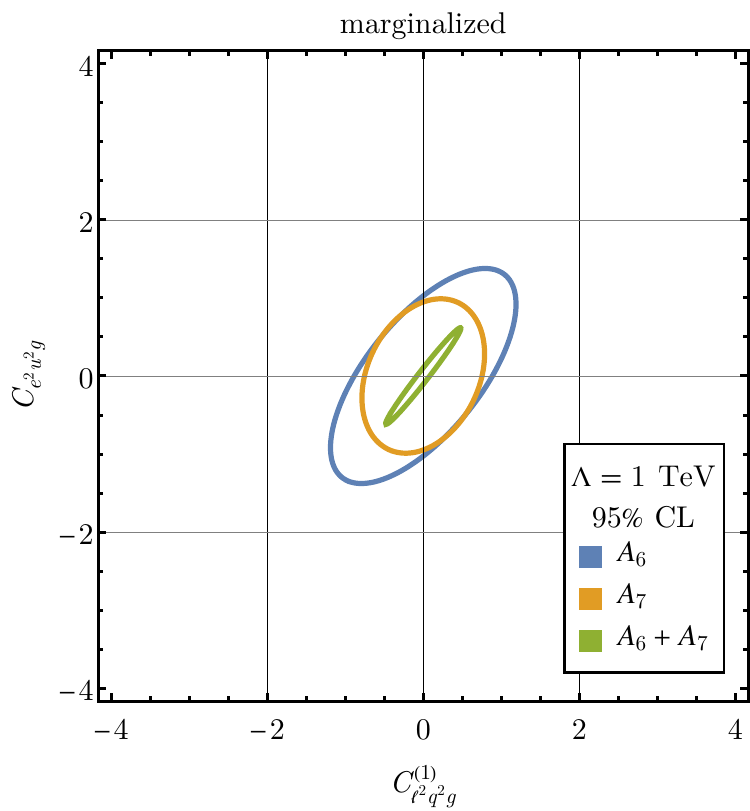}
    \caption{The same as Fig.~\ref{fig:Wil2d} but for $C_{\ell^2 q^2 g}^{(1)}$ and $C_{e^2u^2g}$.}
        \label{fig:UV2d}
\end{figure}

We now study in more detail the correlations between Wilson coefficients that lead to the significant differences between non-marginalized and marginalized fits seen in our 1-dimensional constraints. We present 2-dimensional confidence ellipses in Figures~\ref{fig:Wil2d} and~\ref{fig:UV2d} for a representative selection of Wilson coefficients. Each figure contains both the nonmarginalized fit and the corresponding marginalized ellipses. From these figures, we note the following points.
\begin{itemize}
    \item The $A_6$ and $A_7$ fit results are complementary and yield distinct correlations.
    \item The non-marginalized ellipses yield nearly flat directions with extremely elongated ellipses.
    \item The confidence ellipses projected from the full $7d$ fit do not display flat directions but have significantly weakened constraints.
\end{itemize}


\begin{figure}
    [htbp]
    \centering
    \includegraphics[width=.4875\textwidth]{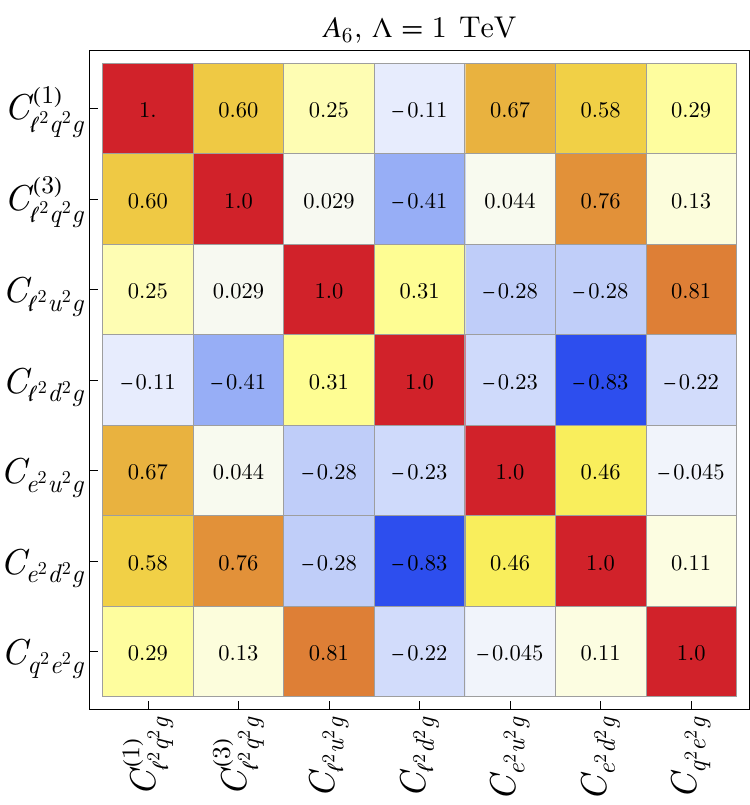}
    \includegraphics[width=.4875\textwidth]{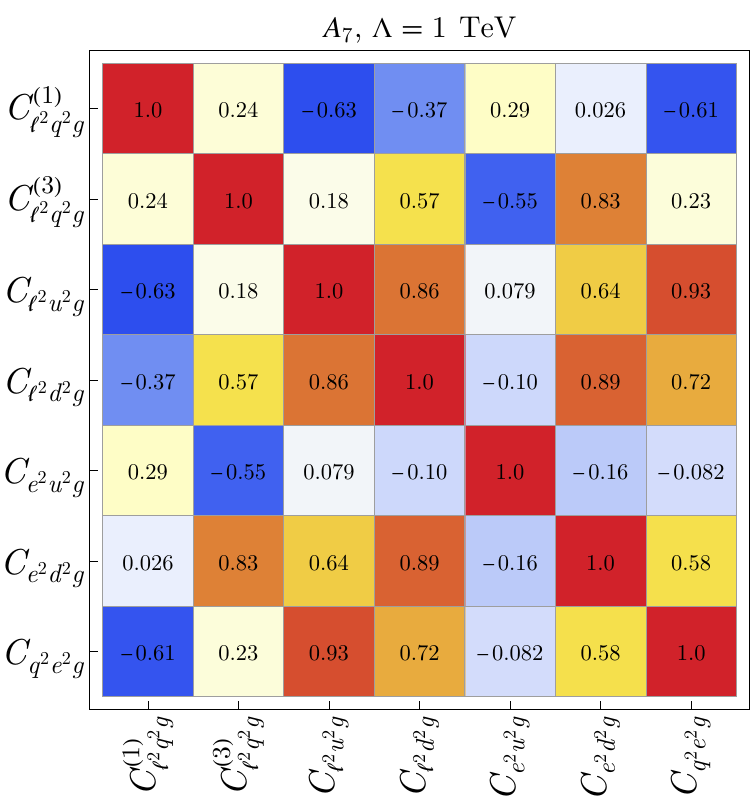}
    \caption{Correlation matrices for the separate $A_6$ and $A_7$ fits.}
    \label{fig:A6_corr_mtr}
\end{figure}
\begin{figure}
    [htbp]
    \centering
    \includegraphics[width=.4875\textwidth]{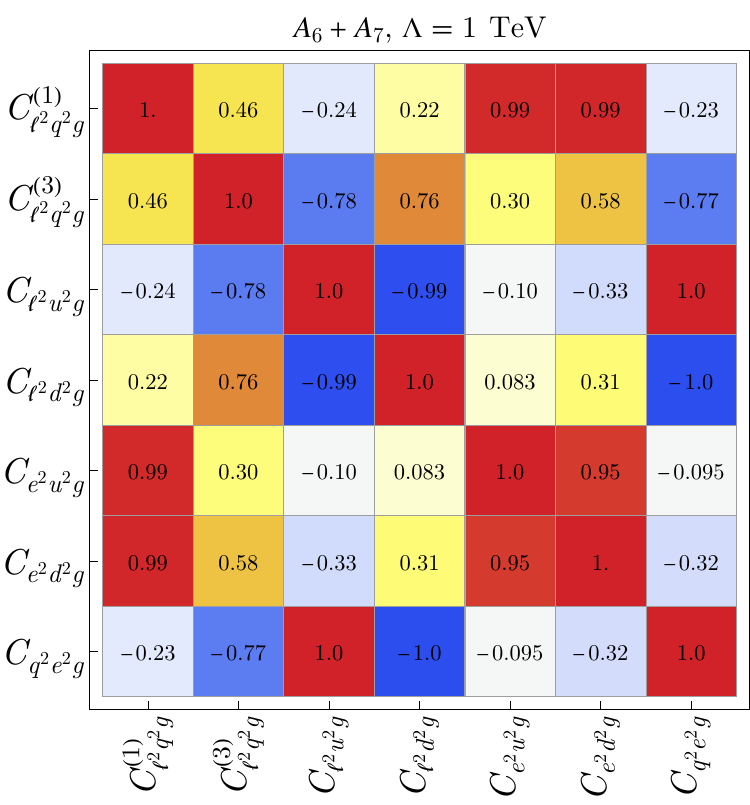}
    \caption{Correlation matrices for the combined $A_6+A_7$ fit.}
    \label{fig:A67_corr_mtr}
\end{figure}

\begin{figure}
    [htbp]
    \centering
    \includegraphics[width=0.4875\textwidth]{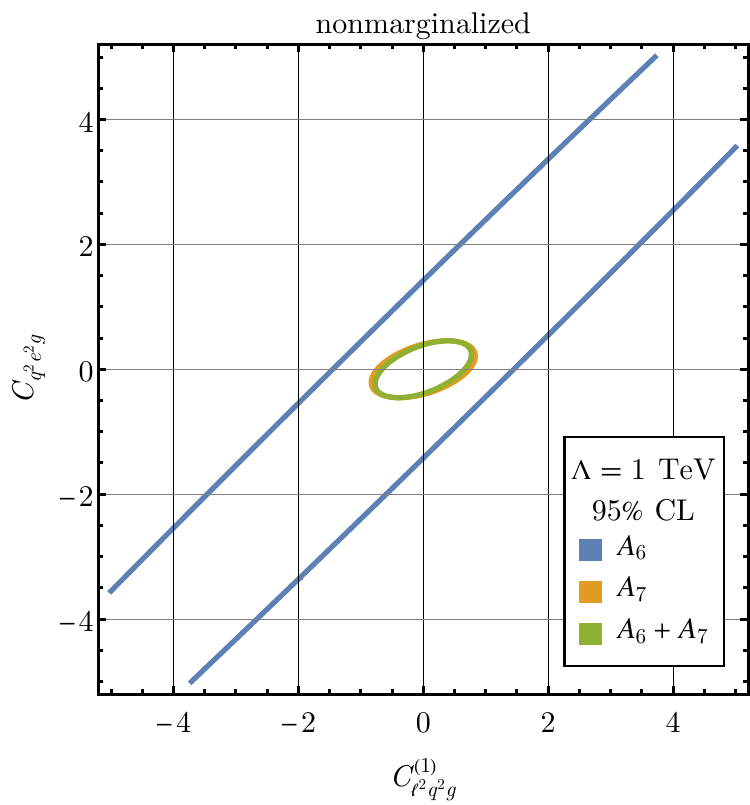}
    \includegraphics[width=0.4875\textwidth]{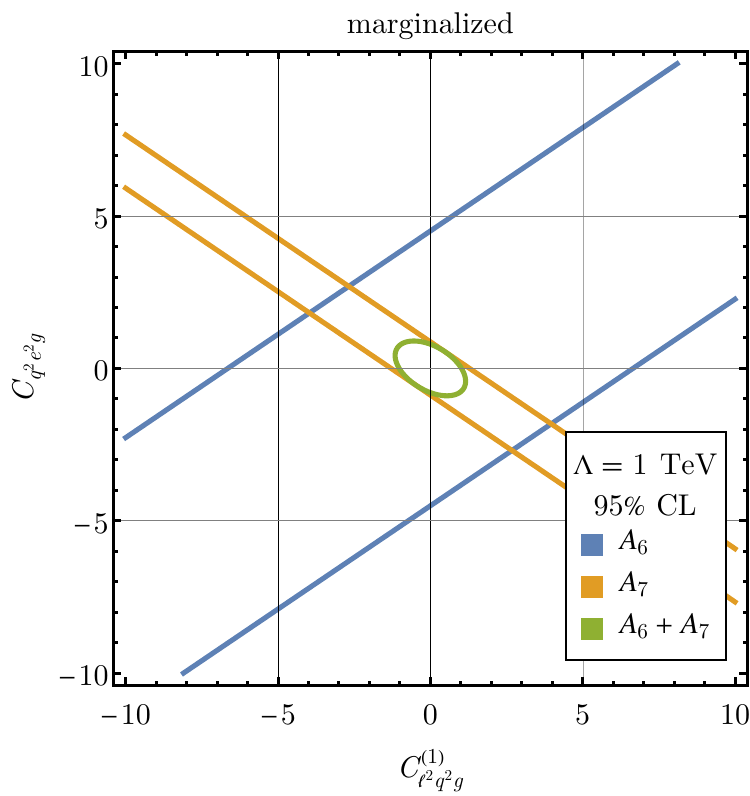}
    \caption{The 95\% confidence ellipses for the Wilson coefficients $C_{\ell^2 q^2g}^{(1)}$ and $C_{q^2e^2g}$ in the $A_6$ and $A_7$ fits, as well as the combined fit, in the nonmarginalized (left) and marginalized (right) cases with the flat directions constraints imposed.}
    \label{fig:fd_ellipse}
\end{figure}

The correlation matrices are presented in Figs.~ \ref{fig:A6_corr_mtr} and~\ref{fig:A67_corr_mtr}. We note large off-diagonal elements, particularly for the correlation matrix for the combined $A_6+A_7$ fit. 
These matrices demonstrate that the $[\bar \ell \gamma^\mu P_X \ell] [\bar q \gamma^\nu P_Y q] G_{\mu\nu}$ sector of the SMEFT suffers from flat directions. We now study these in more detail analytically, to gain some understanding of how these develop. There are two primary subprocesses that contribute to the DY production with an associated jet: pair annihilation, $q\bar q \to g e^-e^+$, and  Compton scattering, $qg \to qe^-e^+$. Each of these processes can be initiated by either an up quark or a down quark. To search for flat directions, we require the SMEFT part of the squared amplitudes to vanish in order to obtain four dependent and three independent Wilson coefficients. We focus on the processes $u\bar u$, $ug$, $d\bar d$, and $dg$ to catch the independent structures; the flat-direction solutions for the other partonic channels are identical to these. Letting $\{C_{\ell^2 q^2 g}^{(1)}, C_{\ell^2 q^2 g}^{(3)}, C_{q^2e^2g}\}$ be the set of independent Wilson coefficients and $\{C_{\ell^2 u^2 g}, C_{\ell^2 d^2 g}, C_{e^2 u^2 g}, C_{e^2d^2g}\}$ the dependent ones, we obtain
\begin{gather}
    C_{\ell^2 u^2 g} = \frac{C_{q^2e^2g} \left(D_Z C_{uuZ}^L C_{\ell \ell Z}^R+D_{\gamma } C_{uu\gamma } C_{\ell \ell \gamma }\right)}{D_Z C_{\ell \ell Z}^L C_{uuZ}^R+D_{\gamma } C_{uu\gamma } C_{\ell
   \ell \gamma }}, \\
    C_{\ell^2 d^2 g} = \frac{C_{q^2e^2g} \left(D_Z C_{ddZ}^L C_{\ell \ell Z}^R+D_{\gamma } C_{dd\gamma } C_{\ell \ell \gamma }\right)}{D_Z C_{ddZ}^R C_{\ell \ell Z}^L+D_{\gamma } C_{dd\gamma } C_{\ell
   \ell \gamma }}, \\
    C_{e^2 u^2 g} = \frac{\left(C_{\ell ^2q^2g}^{\text{(1)}}-C_{\ell ^2q^2g}^{\text{(3)}}\right) \left(D_Z C_{uuZ}^L C_{\ell \ell Z}^L+D_{\gamma } C_{uu\gamma } C_{\ell \ell \gamma }\right)}{D_Z
   C_{uuZ}^R C_{\ell \ell Z}^R+D_{\gamma } C_{uu\gamma } C_{\ell \ell \gamma }}, \\
    C_{e^2 d^2 g} = \frac{\left(C_{\ell ^2q^2g}^{\text{(1)}}+C_{\ell ^2q^2g}^{\text{(3)}}\right) \left(D_Z C_{ddZ}^L C_{\ell \ell Z}^L+D_{\gamma } C_{dd\gamma } C_{\ell \ell \gamma }\right)}{D_Z
   C_{ddZ}^R C_{\ell \ell Z}^R+D_{\gamma } C_{dd\gamma } C_{\ell \ell \gamma }}, 
\end{gather}
where $D_\gamma=1/m_{\ell\ell}{}^2$ and $D_Z=1/(m_{\ell\ell}{}^2-m_Z{}^2)$ are the photon and $Z$-boson propagators. These exact solutions are energy dependent, so any flat direction is not exact but rather approximate. In the high dilepton invariant mass limit, $m_{\ell\ell} \gg m_Z$, these become energy independent:
\begin{gather}
    C_{\ell^2 u^2 g} \approx \frac{C_{q^2e^2g} \left(C_{uuZ}^L C_{\ell \ell Z}^R+C_{uu\gamma } C_{\ell \ell \gamma }\right)}{C_{\ell \ell Z}^L C_{uuZ}^R+C_{uu\gamma } C_{\ell \ell \gamma }}, \\
    C_{\ell^2 d^2 g} \approx \frac{C_{q^2e^2g} \left(C_{ddZ}^L C_{\ell \ell Z}^R+C_{dd\gamma } C_{\ell \ell \gamma }\right)}{C_{ddZ}^R C_{\ell \ell Z}^L+C_{dd\gamma } C_{\ell \ell \gamma }}, \\
    C_{e^2 u^2 g} \approx \frac{\left(C_{\ell ^2q^2g}^{\text{(1)}}-C_{\ell ^2q^2g}^{\text{(3)}}\right) \left(C_{uuZ}^L C_{\ell \ell Z}^L+C_{uu\gamma } C_{\ell \ell \gamma }\right)}{C_{uuZ}^R C_{\ell \ell
   Z}^R+C_{uu\gamma } C_{\ell \ell \gamma }}, \\
    C_{e^2 d^2 g} \approx \frac{\left(C_{\ell ^2q^2g}^{\text{(1)}}+C_{\ell ^2q^2g}^{\text{(3)}}\right) \left(C_{ddZ}^L C_{\ell \ell Z}^L+C_{dd\gamma } C_{\ell \ell \gamma }\right)}{C_{ddZ}^R C_{\ell \ell
   Z}^R+C_{dd\gamma } C_{\ell \ell \gamma }}.
\end{gather}
With our input scheme, we obtain the following approximate numerical conditions for the SMEFT corrections to vanish:
\begin{gather}
    C_{\ell^2 u^2 g} \approx \frac{C_{q^2e^2g}}{2}, \\
    C_{\ell^2 d^2 g} \approx -C_{q^2e^2g}, \\
    C_{e^2 u^2 g} \approx -\frac{\left(2 s_W{}^2-3\right) \left(C_{\ell ^2q^2g}^{\text{(1)}}-C_{\ell ^2q^2g}^{\text{(3)}}\right)}{8 s_W{}^2} = 1.37246 \left(C_{\ell ^2q^2g}^{\text{(1)}}-C_{\ell ^2q^2g}^{\text{(3)}}\right), \\
    C_{e^2 d^2 g} \approx -\frac{\left(4 s_W{}^2-3\right) \left(C_{\ell ^2q^2g}^{\text{(1)}}+C_{\ell ^2q^2g}^{\text{(3)}}\right)}{4 s_W{}^2} = 2.24493 \left(C_{\ell ^2q^2g}^{\text{(1)}}+C_{\ell ^2q^2g}^{\text{(3)}}\right).
\end{gather}
We then use these relations in the squared amplitudes and perform a 3-parameter fit. 
In the $A_6$ and $A_7$ fits, we obtain perfect correlations even with only approximate flat directions. The lowest $m_{\ell\ell}$ bin is [300, 360] GeV, which is several times the $Z$ mass. The combined fit yields strong correlations but not at the extremes. An example of the confidence ellipses are shown in Fig.~\ref{fig:fd_ellipse}. The importance of measuring both $A_6$ and $A_7$ is clear from these plots. A single observable cannot remove the flat directions. Similar flat directions in the high-energy Drell-Yan process have been observed previously~\cite{Boughezal:2020uwq}. Their appearance highlights the need for a broad-based experimental program with input from multiple experiments to fully probe the SMEFT parameter space.

\section{Conclusion}
\label{sec:conc}

We have proposed here the naive $T$-odd Collins Soper angular moments
in the neutral-current Drell-Yan process at high transverse momentum and invariant mass as a probe of heavy new $CP$-odd physics. Our analysis has used the framework of the Standard Model Effective Field Theory (SMEFT) to provide a model-independent parametrization for heavy new physics. Experimental studies of these
quantities at the LHC have focused on the $Z$-peak region. We have
shown that extending these measurements beyond the $Z$-peak can probe
unexplored directions in the SMEFT parameter space. The expected
luminosity at a future high-luminosity LHC will allow the 1-2 TeV scale to be
probed with a combined fit of the $A_{6-7}$ moments. We encourage our
experimental colleagues to perform these measurements with future data.

\medskip

\noindent
\textbf{Acknowledgments}: FP is supported by the U.S. Department of Energy, Office of High Energy Physics, under contract No. DE-SC0010143. K\c{S} is supported by the Kennesaw State University Office of Research Postdoctoral Fellowship Program. This research was supported in part through the computational resources and staff contributions provided for the Quest high performance computing facility at Northwestern University which is jointly supported by the Office of the Provost, the Office for Research, and Northwestern University Information Technology.

\noindent
\textbf{Data availability}: The data and code associated with this article are available at Ref.~\cite{kaan_simsek_2025_17905989}.

\bibliography{refs}

\end{document}